\documentclass[twocolumn]{aastex631}

\hypersetup{linkcolor=red,citecolor=blue,filecolor=cyan,urlcolor=magenta}

\def\lessim{\mathrel{\hbox{\rlap{\hbox{\lower4pt\hbox{$\sim$}}}\hbox{$<$}}}}
\def\grtsim{\mathrel{\hbox{\rlap{\hbox{\lower4pt\hbox{$\sim$}}}\hbox{$>$}}}}

\usepackage{CJK}

\shorttitle{MMRD}
\shortauthors{Shafter et al.}

\graphicspath{{./}{figures/}}

\begin{document}

\title{Concerning the Verity of the MMRD Relation for Novae}

\correspondingauthor{A. W. Shafter}
\email{ashafter@sdsu.edu}

\author[0000-0002-1276-1486]{Allen W. Shafter}
\affiliation{Department of Astronomy, San Diego State University, San Diego, CA 92182, USA}

\author[0000-0003-4735-9128]{J. Grace Clark}
\affiliation{Department of Astronomy, San Diego State University, San Diego, CA 92182, USA}

\author[0000-0002-0835-225X]{Kamil Hornoch}
\affiliation{Astronomical Institute of the Czech Academy of Sciences, Fri\v{c}ova 298, CZ-251 65 Ond\v{r}ejov, Czech Republic}

\begin{abstract}

It has long been claimed that novae reaching the highest luminosity at the peak of their eruptions appear to fade the
fastest from maximum light.
The relationship between peak brightness and fade rate is known as the Maximum-Magnitude, Rate-of-Decline (MMRD) relation. Lightcurve parameters for the most recent sample of M31 recurrent novae are presented and used to buttress the case
that the observed MMRD relation can be explained as a consequence of observational selection effects coupled with expectations from standard nova models.

\end{abstract}


\section{Introduction}

\citet{1945PASP...57...69M} was first to argue that the peak luminosity of a classical nova was correlated with its rate of decline from maximum light.
Over the years, the correlation has come to be known as the Maximum-Magnitude versus Rate-of-Decline (MMRD) relation, and
has been calibrated many times, both in the Galaxy and in M31 \citep[e.g.,][]{1985ApJ...292...90C,1989AJ.....97.1622C,2000AJ....120.2007D,2011ApJ...734...12S,2018MNRAS.476.4162O}.
Despite its long history,
the verity of the MMRD has been called into question in recent years.

In a sample of M31 novae, \citet{2011ApJ...735...94K} found several systems that
were fainter for a given decline rate than predicted by the MMRD.
They considered the
possibility that these objects might be recurrent novae (RNe),
but noted that most
had spectroscopic types (Fe~II) that were inconsistent with
that interpretation.

\begin{figure*}
\plotone{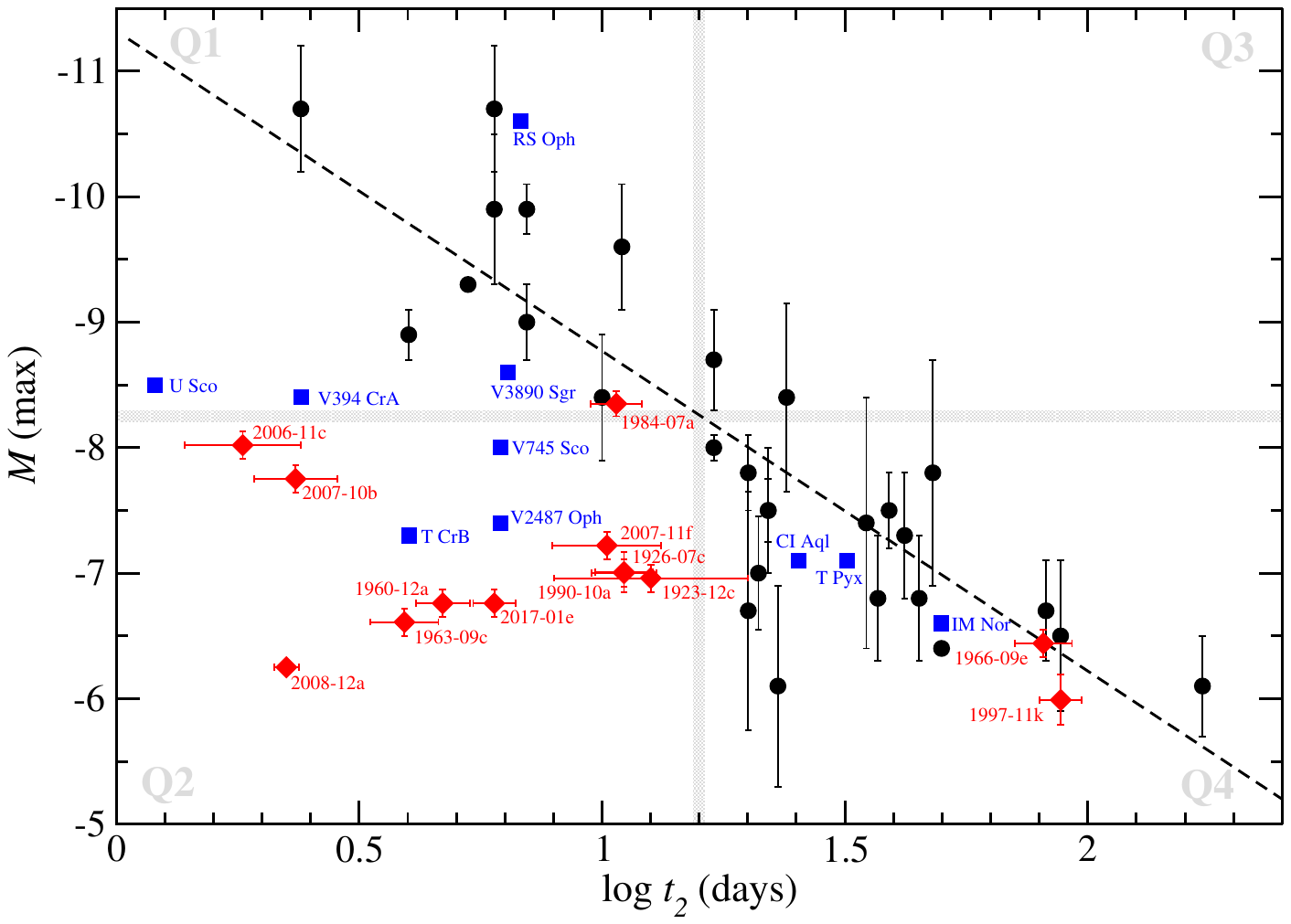}
\caption{The MMRD plane divided into four quadrants: Q1 -- Q4. The filled black circles show data for Galactic novae from \citet{2000AJ....120.2007D}. The dashed line is the best linear fit to these data. The blue squares show Galactic
RNe from \citet{2010ApJS..187..275S}, while the red diamonds show our updated M31 RN sample. Most classical novae discovered in routine nova patrols fall either in Q1 or Q4. Known RNe on the other hand fall almost exclusively in Q2, while Q3 is almost devoid of novae.
}
\label{fig:f1}
\end{figure*}

Subsequent to the \citet{2011ApJ...735...94K} study,
\citet{2018MNRAS.476.4162O}
compiled an extensive database for Galactic novae showing
that the MMRD relation was generally followed and
remained ``a useful tool for statistical analyses".
Shortly thereafter, \citet{2018MNRAS.481.3033S} recalibrated
the Galactic MMRD based on {\it Gaia\/} distances concluding that
the MMRD relation was plagued by considerable scatter with a fit too poor to be usable for distance determinations.

In this note, we argue that the MMRD relation can be best understood as a combination of observational selection effects coupled with standard predictions from nova theory.

\section{The Nature of the MMRD}

To understand the placement of novae in the $M$ (max) -- log~$t_2$ plane (hereafter the MMRD plane) we first consider how
the ignition mass -- the accreted mass required to trigger a thermonuclear runaway (TNR) on the white dwarf -- depends on properties of the progenitor binary. To first order, $M_\mathrm{ign}$ depends only on the
pressure at the base of the accreted layer, which is a function of the WD mass.
However, the temperature of the accreted layer is also important, and it is strongly affected
by the rate of accretion onto the white dwarf. Thus, as first considered by \cite{1982ApJ...253..798N} and explored by many groups since \citep[e.g.,][]{2005ApJ...628..395T,2013ApJ...777..136W,2014ApJ...793..136K},
the mass necessary to trigger a TNR depends on both the WD mass and the rate of accretion onto its
surface.

For illustrative purposes, we consider in Figure~\ref{fig:f1} the MMRD relation for the sample of Galactic novae studied by \citet{2000AJ....120.2007D}. In addition, we show separately the known Galactic RNe from \citet{2010ApJS..187..275S}, along with the most recent M31 RN sample presented here together for the first time.
Despite some scatter, the Downes \& Duerbeck nova sample follows the best-fitting MMRD relation (dashed line) quite well.
However, the RNe fall consistently below the MMRD relation.

To explore the observed properties of the MMRD in detail,
it is useful to divide the MMRD plane into four quadrants: (Q1) an upper left quadrant consisting of relatively fast and bright novae, (Q2) a lower left quadrant that includes fast and faint novae, (Q3) an upper right quadrant where slowly evolving luminous novae should lie, and finally (Q4) a lower right quadrant where slowly evolving and faint novae are found.

Given a population of nova progenitors with a range of WD masses and accretion rates, one can imagine systems occupying all quadrants of the MMRD plane. However, we argue below that
the observed MMRD relation arises because two of the quadrants, the second and especially the third, are selected against.

In the case of the
second quadrant, the low luminosity and fast evolution (short $t_2$) suggests a weak TNR and a relatively small ejected (and ignition) mass. Models show that novae with these properties arise from systems with high mass white dwarfs accreting at high rates. The small ignition masses and high accretion rates produce novae with the shortest recurrence times
\citep[e.g., see][their figure 6]{2014ApJ...793..136K}. Thus, it is not surprising that the known RNe are found in the lower left quadrant of the MMRD plane. Considering their short recurrence times, it is reasonable to wonder why this quadrant of the MMRD plane is not more heavily populated. The answer lies in the fact that
faint and fast novae are strongly selected against in typical nova surveys, most of which have relatively bright limiting magnitudes and coarse temporal coverage.

In the third quadrant of the MMRD plane we expect to find novae that are luminous and slowly evolving. The slow evolution suggests a massive ejecta (and ignition mass), while the high luminosity implies a strong TNR. Models suggest that the progenitors of such novae contain relatively low mass white dwarfs accreting at low rates. The slow accumulation of matter on the white dwarf and a high ignition mass results in both a strong TNR and a very long recurrence time. Such systems are the polar opposite of the RNe. Although they should have bright eruptions, the recurrence times are expected to be exceedingly long. Thus, systems in the upper right quadrant of the MMRD are expected to erupt extremely rarely, in agreement with observation.

Consistent with the nature of the MMRD relation, most novae fall into either the first or the fourth quadrants. Novae in the first quadrant presumably arise from novae with high mass white dwarfs accreting at relatively low rates, while novae in the fourth quadrant likely have progenitors that contain relatively low mass white dwarfs accreting at relatively high rates. In the first case, the high mass WD and the slow accretion will result in an accreted layer that is highly degenerate at the time when the TNR ensues. Such novae should appear relatively bright and evolve quickly. Conversely, the rapidly accreting low mass WD systems will be characterized by less degenerate accreted layers and a weaker TNR resulting in less luminous novae with a generally slower evolution.

In summary, the MMRD emerges as a result of observational selection against faint and fast (recurrent) novae coupled with a dearth of eruptions from systems with extremely long recurrence times.

\bibliography{novarefs}{}
\bibliographystyle{aasjournal}

\appendix

\startlongtable
\begin{deluxetable*}{lclcc}
\tablenum{A1}
\tablecolumns{5}
\tablecaption{MMRD Lightcurve Parameters (Figure 1)}
\tablehead{\colhead{Nova} & \colhead{$t_2$ (d)} & \colhead{$M$(max)\tablenotemark{a}} & \colhead{Filter} & \colhead{Note\tablenotemark{b}}}
\startdata
\cutinhead{Galactic Nova Sample}
V500 Aql     &  17.0       &$ -8.70\pm 0.40$& pg&1 \cr
V603 Aql     &   4.0       &$ -8.90\pm 0.20$&$V$&1 \cr
V1229 Aql    &  20.0       &$ -6.70\pm 0.95$&$V$&1 \cr
T Aur        &  45.0       &$ -6.80\pm 0.50$& pg&1 \cr
V842 Cen     &  35.0       &$ -7.40\pm 1.00$&$V$&1 \cr
V450 Cyg     &  88.0       &$ -6.50\pm 0.60$&$V$&1 \cr
V476 Cyg     &   6.0       &$ -9.90\pm 0.60$&$V$&1 \cr
V1500 Vyg    &   2.4       &$-10.70\pm 0.50$&$V$&1 \cr
V1819 Cyg    &  37.0       &$ -6.80\pm 0.50$&$V$&1 \cr
V1974 Cyg    &  17.0       &$ -8.00\pm 0.10$&$V$&1 \cr
HR Del       & 172.0       &$ -6.10\pm 0.40$&$V$&1 \cr
DQ Her       &  39.0       &$ -7.50\pm 0.30$&$V$&1 \cr
V446 Her     &   7.0       &$ -9.90\pm 0.20$& pg&1 \cr
V533 Her     &  22.0       &$ -7.50\pm 0.50$&$V$&1 \cr
CP Lac       &   5.3       &$ -9.30        $&$V$&1 \cr
DK Lac       &  11.0       &$ -9.60\pm 0.50$&$V$&1 \cr
GK Per       &   7.0       &$ -9.00\pm 0.30$&$V$&1 \cr
RR Pic       &  20.0       &$ -7.80\pm 0.30$&$V$&1 \cr
CP Pup       &   6.0       &$-10.70\pm 0.50$&$V$&1 \cr
V351 Pup     &  10.0       &$ -8.40\pm 0.50$&$V$&1 \cr
FH Ser       &  42.0       &$ -7.30\pm 0.50$&$V$&1 \cr
XX Tau       &  24.0       &$ -8.40\pm 0.75$& pg&1 \cr
RW UMi       &  48.0       &$ -7.80\pm 0.90$& pg&1 \cr
LV Vul       &  21.0       &$ -7.00\pm 0.45$&$V$&1 \cr
NQ Vul       &  23.0       &$ -6.10\pm 0.80$&$V$&1 \cr
PW Vul       &  82.0       &$ -6.70\pm 0.40$&$V$&1 \cr
QU Vul       &  22.0       &$ -7.50\pm 0.20$&$V$&1 \cr
QV Vul       &  50.0       &$ -6.40        $&$V$&1 \cr
\cutinhead{Galactic Recurrent Novae}
T Pyx        &  32.0       &$ -7.10        $&$V$&2 \cr
IM Nor       &  50.0       &$ -6.60        $&$V$&2 \cr
CI Aql       &  25.4       &$ -7.10        $&$V$&2 \cr
V2487 Oph    &   6.2       &$ -7.40        $&$V$&2 \cr
U Sco        &   1.2       &$ -8.50        $&$V$&2 \cr
V394 CrA     &   2.4       &$ -8.40        $&$V$&2 \cr
T CrB        &   4.0       &$ -7.30        $&$V$&2 \cr
RS Oph       &   6.8       &$-10.60        $&$V$&2 \cr
V745 Sco     &   6.2       &$ -8.00        $&$V$&2 \cr
V3890 Sgr    &   6.4       &$ -8.60        $&$V$&2 \cr
\cutinhead{M31 Recurrent Novae}
M31N 1923-12c&$ 12.6\pm5.8$&$ -6.96\pm 0.11$&$R$&3 \cr
M31N 1926-07c&$ 11.1\pm1.5$&$ -7.01\pm 0.16$&$R$&3 \cr
M31N 1960-12a&$  4.7\pm0.6$&$ -6.76\pm 0.11$&$R$&3 \cr
M31N 1963-09c&$  3.9\pm0.6$&$ -6.61\pm 0.11$&$R$&3 \cr
M31N 1966-09e&$81.0\pm11.0$&$ -6.44\pm 0.11$&$R$&3 \cr
M31N 1984-07a&$ 10.7\pm1.3$&$ -8.35\pm 0.10$&$R$&3 \cr
M31N 1990-10a&$ 11.1\pm1.7$&$ -7.00\pm 0.11$&$R$&3 \cr
M31N 1997-11k&$ 88.0\pm9.0$&$ -5.99\pm 0.20$&$R$&3 \cr
M31N 2006-11c&$  1.8\pm0.5$&$ -8.02\pm 0.11$&$R$&3 \cr
M31N 2007-10b&$  2.3\pm0.5$&$ -7.75\pm 0.11$&$R$&3 \cr
M31N 2007-11f&$ 10.2\pm2.6$&$ -7.22\pm 0.11$&$R$&3 \cr
M31N 2008-12a&$  2.2\pm0.1$&$ -6.25\pm 0.04$&$R$&3 \cr
M31N 2017-01e&$  6.0\pm0.6$&$ -6.76\pm 0.11$&$R$&3 \cr
\enddata
\tablenotetext{a}{$M$(max) based on $m-M=24.42$ \citep[][]{2013AJ....146...86T} and $A_R=0.14$ \citep[][]{2011ApJ...737..103S} for M31.}
\tablenotetext{b}{(1) \citet{2000AJ....120.2007D}; (2) \citet{2010ApJS..187..275S}; (3) This work.}
\end{deluxetable*}

\end{document}